\documentclass{aa}     
\usepackage{graphics}
\def\cm{\,{\rm cm}}
\def\cc{\,{\rm cm^{-3}}}
\def\cm2{\,{\rm cm^{-2}}}

\def\kms{\,{\rm {km\,s^{-1}}}}
\def\kkms{\,{\rm {K\,km\,s^{-1}}}}

\def\h2{\,{\rm H_{2}}}
\def\13co{\,{\rm ^{13}CO}}
\def\co{\,{\rm ^{12}CO}}
\def\pci{\,{\rm ^{3}P_{1}-^{3}P_{0}\,[CI]}}

\def\aua{A\&A }
\def\auas{A\&AS }
\def\apj{ApJ }
\def\pasj{PASJ }
\def\aj{AJ }
\def\apjs{ApJS }
\def\apjl{ApJL }

\begin{document}
 
   \thesaurus{03         
              (11.09.1 M\,31; 
	       11.09.4; 
               09.09.1 D\,478
               09.13.2,
               13.19.1)
             }
   \title{Detection of neutral carbon in the M~31 dark cloud D~478}
 
   \subtitle{}
 
   \author{F.P. Israel
          \inst{1}
           and R.P.J. Tilanus
	  \inst{2,3}
           and F. Baas
          \inst{1,2}
}
 
   \offprints{F.P. Israel}
 
  \institute{Sterrewacht Leiden, P.O. Box 9513, NL 2300 RA Leiden,
             The Netherlands
  \and       Joint Astronomy Centre, 660 N. A'ohoku Pl., Hilo,
             Hawaii, 96720, USA
  \and       Netherlands Foundation for Research in Astronomy,
             P.O. Box 2, NL 7990 AA Dwingeloo, The Netherlands}
   \date{Received 17 April 1998; accepted 7 July 1998}
 
   \maketitle
 
   \begin{abstract}

Emission from the $\pci$ transition at 492 GHz has been detected towards the
dark cloud D\,478 in the Local Group galaxy M~31. Using literature
detections of the lower $\co$ and $\13co$ transitions, models for the gas 
distribution in D~478 are discussed. The observed CO and C line ratios can 
be explained by two-component models (high density cores and low-density 
envelopes); single-density models appear less likely. The models indicate 
kinetic temperatures of the order of T$_{\rm k} \approx$ 10 K. The 
beam-averaged column density of neutral carbon is 0.3--0.8 times that of CO,
whereas the total carbon to hydrogen ratio N$_{\rm C}$/N$_{\rm H}$ is
5--3 $\times$ 10$^{-4}$. The resulting CO to $\h2$ conversion factor $X$ 
is about half that of the Solar Neighbourhood. With temperatures of about 
10 K and projected mass-densities of 5--10 M$_{\odot}$ pc$^{-2}$ there 
appears to be no need to invoke the presence of very cold and very massive 
clouds. Rather, D~478 appears to be comparable to Milky Way dark cloud 
complexes at the higher metallicity expected from its central location in 
M~31. In particular, several similarities between D~478 and the Galactic
Taurus-Auriga dark cloud complex may be noted.


\end{abstract}
 
\section{Introduction}

Several molecular cloud complexes have been detected in the spiral
arms of the Local Group galaxy M~31 by their $\co$ $J$=1--0 emission
(Blitz 1985; Ryden $\&$ Stark 1986; Vogel et al. 1987;
Casoli et al. 1987; Nakano et al. 1987; Lada et al. 1988;
Casoli $\&$ Combes 1988; Berkhuijsen et al. 1993; Wilson
$\&$ Rudolph 1993; Loinard et al. 1996b). Most appear to be giant
molecular cloud complexes similar to those in the disk of the Milky
Way (Vogel et al 1987; Lada et al. 1988; Wilson $\&$ Rudolph 1993). 
However, Blitz (1985) noted velocity widths systematically higher than 
expected on the basis of the observed integrated line intensities. He 
proposed that the observed CO emission is from both GMC's and small
clouds, and dominated by the contribution from the latter. Maps of a
northeastern spiral arm of M~31 by Casoli et al. (1987) and Casoli
$\&$ Combes (1988) and a southwestern spiral arm by Kutner et al.
(1990) provide evidence for a dual population: in
addition to major complexes that have the characteristics of `normal'
GMC's, additional CO emission appears to originate from numerous small
clouds. Unbiased CO surveys along the minor and major axes of M~31
have been carried out by Sandqvist et al. (1989) and
Loinard et al. (1995) respectively. The former found
relatively strong CO emission associated with M~31 spiral arm `4' and
weaker CO emission from the outer warp. The latter found CO emission
in the majority of positions sampled, in a number of cases without a
counterpart dust cloud. They also found generally low $\co$
$J$=2--1/$J$=1--0 ratios which they interpreted as arising from cold
($T_{\rm k} <$ 5 K) molecular clouds at low density ($n_{\rm H2}
\approx$ 100 cm$^{-3}$).

In addition, they observed several dust clouds (Hodge 1980) in the
inner parts of M~31 (Allen $\&$ Lequeux 1993; Loinard et al. 1996a). 
Two of these, D~268 and D~478 were observed in the
lower two transitions of both $\co$ and $\13co$ and subsequently
modelled in detail by Allen et al. (1995). They conclude that the
$\co$ emission is dominated by a very cold, low-density gas, while the
$\13co$ emission comes largely from higher-density clumps inside the
clouds. In an attempt to further investigate the unusual physical
conditions of these molecular clouds, we have observed the strongest
of these two objects, D~478, in the neutral carbon line. While this
work was in progress, an interferometric map of the $\co$ $J$=1--0
distribution, and the $\co$ $J$=3--2 profile of D~478 have also become
available (Loinard $\&$ Allen 1998).

\section{Observations}

The observations were carried out with the 15m James Clerk Maxwell
Telescope (JCMT) on Mauna Kea (Hawaii) \footnote{The James Clerk 
Maxwell Telescope is operated by The Joint Astronomy Centre on behalf 
of the Particle Physics and Astronomy Research Council of the United 
Kingdom, the Netherlands Organisation for Scientific Research, and 
the National Research Council of Canada.}. The $\pci$ transition at 
$\nu$ = 492.161 GHz was observed towards D~478 for a total integration 
time of 100 minutes (on+off) on November 29, 1996.  At the observing 
frequency the resolution was 10.2$''$ (HPBW) compared to a pointing 
accuracy better than 2$''$ r.m.s. Weather conditions were excellent, 
resulting in an overall system temperature including sky of $T_{\rm sys}$ 
= 1365 K.  For a backend, we used the DAS digital autocorrelator system 
in a band of 250 MHz. The resulting spectrum was binned to a velocity
resolution of 4.6 $\kms$. At this velocity resolution, the
r.m.s. noise is about 14 mK in $T_{\rm A}^{*}$. We applied a linear
baseline correction only and scaled the spectrum to a main-beam
brightness temperature, $T_{\rm mb}$ = $T_{\rm A}^{*}$/$\eta _{\rm
mb}$ using $\eta _{\rm mb}$ = 0.53.  Line parameters were determined
by gaussian fitting; the results are given in Table 1. The spectrum is
shown in Fig. 1.

\begin{figure}
\resizebox{11cm}{!}{\rotatebox{270}{\includegraphics*{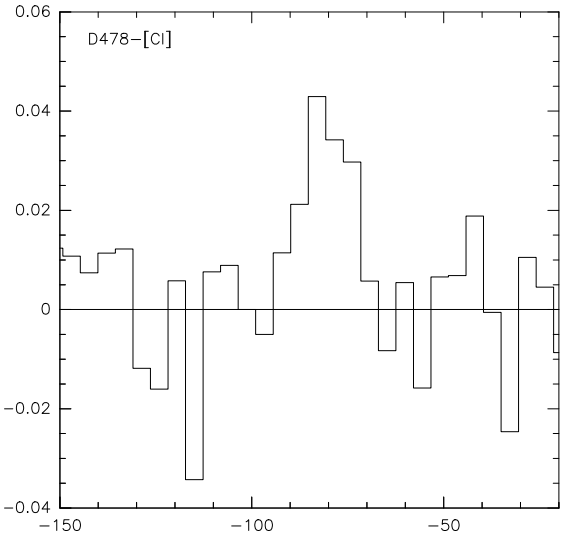}}}
\caption[]{[CI] spectrum observed towards D~478 in M~31. The vertical 
scale is $T_{\rm A}^{*}$ = 0.53 $T_{\rm mb}$ in K; the horizontal scale is
velocity $V_{\rm LSR}$ in $\kms$.  }
\end{figure}

\begin{table}
\caption[]{[CI] and CO in D~478}
\begin{flushleft}
\begin{tabular}{llcccc}
\hline
\noalign{\smallskip}
Transition		& $T_{\rm mb}$ & $\Delta V$ & $\int T_{\rm mb}$d$V$ & $V_{\rm hel}$ \\
			& (mK)	       & $\kms$     & K $\kms$ 		    & $\kms$ \\
\noalign{\smallskip}
\hline
\noalign{\smallskip}
$\pci$			& 85$\,\pm\,$12     & 15$\,\pm\,$2.5 & 1.32$\,\pm\,$0.18 	    & -85.1 \\
\noalign{\smallskip}
$\co$ $J$=2--1$^{a}$ 	& 310	       & 19.5	    & 6.44$\,\pm\,$0.15    	    & -85.2 \\
\noalign{\smallskip}
\hline
\end{tabular}
\end{flushleft}
Notes: a. Values in a 13$''$ beam (J. Lequeux, private communication)
\end{table}

\section{Results and analysis}

\subsection{[CI]/CO ratio}

The 10$''$ [CI] profile in Fig. 1 is very similar to those of 
$\co$ $J$=2--1 and $J$=3--2 at 15$''$ (Loinard $\&$ Allen 1998),
suggesting similar spatial distributions of CO and C. The effect of 
resolution may be estimated by comparing the $\co$ $J$=2--1 at a 
full resolution of 13$''$ (kindly supplied by J. Lequeux, see Table 1) 
with the $\co$ result convolved to 23$''$ given by Allen et al. (1995). 
In amplitude, the full resolution profile is a factor 1.35 higher and 
in area it is a factor of 1.18 greater. We therefore feel confident in 
comparing the [CI]/$\co$ $J$=2--1 ratios resulting from our Table 1 to the 
CO ratios implicit in Allen et al. (1995 -- their Table 1). It then follows 
that the [CI] 492 GHz line is similar in strength to the $\13co$ $J$=1--0 
line, and considerably stronger than the $\13co$ $J$=2--1 line. 
Specifically, we find [CI]/$\13co$(2--1) = 2.0$\,\pm\,$0.5. This is 
similar to the ratios 0.8 -- 1.7 found at various translucent regions 
in the dark Galactic cloud L~183 (Stark et al. 1996), somewhat higher 
than the corresponding range 0.9 -- 1.3 found in the Galactic dark cloud 
TMC-1 (Schilke et al. 1995) and rather higher than the ratio of 0.8 
characterizing the Orion Bar region (Tauber et al. 1995). For this reason, 
we decided to also include translucent conditions, characterized by low 
CO and C column densities, in our analysis in addition to two-component 
models of the type considered by Allen et al. (1995), or the extremely 
low temperature model by Loinard $\&$ Allen (1998).

\begin{table*}
\caption[]{Model parameters and line ratios in D~478}
\begin{flushleft}
\begin{tabular}{lcccccccc}
\hline
\noalign{\smallskip} 
	& \multicolumn{3}{c}{Single-Component Models}  & 
                 \multicolumn{4}{c}{Two-Component Models} & {\bf Observed }\\
\noalign{\smallskip}
\hline
\noalign{\smallskip}
$T_{\rm kin}$ \hfill (K)		     &  6    &  10   &  14   & 6, 6	 & 6, 10    & 10/10	& 10/14     & \\
Weight		     			     & ---   & ---   & ---   & 1/2, 1	 & 1, 1     & 1/6, 1	& 1/4, 1    & \\
n($\h2)$      \hfill        $(\cc)$               & 5000  & 1500  & 750   & \multicolumn{4}{c}{3000, 100}                & \\
$N(\13co)$/d$V$  $(10^{14} \cm2/\kms)$ &  4    &  7    &  8    & 20, 2     & 20, 2    & 60, 6     & 52, 5     & \\
$N(\co)$/d$V$   $(10^{16} \cm2/\kms)$ & 0.8   & 1.3   & 1.7   & \multicolumn{4}{c}{72, 7.2}                  & \\
$N(\rm C)$/d$V$  \hfill $(10^{16} \cm2/\kms)$ &  6    & 2.5   &  2    & 120, 160  & 30, 105  & 15, 80    & 5, 70     & \\
\noalign{\smallskip}
\hline
\noalign{\smallskip}
$\co$ (2--1)/(1--0)   	  		& 0.46  & 0.46  & 0.46  & 0.40 	    & 0.46 	& 0.45	    & 0.48 	& {\bf 0.42$\,\pm\,$0.08}$^{a}$ \\
$\co$ (3--2)/(2--1)   			&{\it 0.12}&{\it 0.21}& 0.26 & 0.32 & 0.30 & 0.28 & 0.32 &{\bf 0.33$\,\pm\,$0.10}$^{b}$ \\
$\13co$ (2--1)/(1--0) 			& 0.40  & 0.40  & 0.40  & 0.34      & 0.35	& 0.48 	    & 0.50 	& {\bf 0.45$\,\pm\,$0.15}$^{a}$ \\
$\co/\13co$ (1--0)    			& 7.4   & 7.8   & 7.8   & 7.4       & 7.4       & 8.5  	    & 8.7 	& {\bf 8.8$\,\pm\,$1.8}$^{a}$ \\
$\co/\13co$ (2--1)    			& 8.5   & 8.7   & 9.1   & 8.4       & 9.7       & 7.5  	    & 8.5 	& {\bf 8.4$\,\pm\,$1.7}$^{a}$ \\
CI/$\co$ (2--1)       			& \multicolumn{7}{c}{0.24} 						& {\bf 0.24$\,\pm\,$0.05}$^{c}$ \\
\noalign{\smallskip}
\hline
\end{tabular}
\end{flushleft}
Notes: a. Derived from Allen et al. (1995); b. Derived from Loinard $\&$ Allen
(1998); c. This Paper.
\end{table*}

\subsection{Model calculations}

Our approach is the following. We first use radiative transfer models
to explore the parameter space allowed by the observed line ratios.
We then use the [CI] observations and chemical model results (van Dishoeck 
$\&$ Black 1988; see also van Dishoeck 1998) to further constrain the 
parameters thus found.

\subsubsection{Radiative transfer modelling}

In order to constrain the range of possible cloud parameters consistent 
with the observed line ratios, we have performed statistical equilibrium 
calculations to determine the population distribution over the ground 
triplet levels of [CI] and the lower $\co$  and $\13co$ rotational levels, 
using an escape probability method for the radiative transfer (cf. Jansen 
1995; Jansen et al. 1994). These calculations
include collisional excitation and de-excitation as well as spontaneous
and stimulated radiative transfer. The equations have been solved to
find the level populations for a range of carbon and carbon monoxide
column densities $N(\co)$, $N(\13co)$ and $N(C)$, kinetic temperatures 
$T_{\rm kin}$ and $\h2$ volume densities $n(\h2)$. We included a cosmic 
background radiation field of 2.75 K. and an incident radiation field
$I_{\rm UV}$ $\approx$ 0.5 ($I_{\rm UV}$ corresponding to
$I_{\rm 1000}$ = 4.5 $\times$ 10$^{-8}$ photons s$^{-1}$ cm$^{-2}$).

As we have four independent CO line ratios, single-component CO models are, 
in principle, fully constrained by the four adjustable parameters. 
In practice, relatively large observational uncertainties allow only 
fitting of three parameters as a function of the fourth, for which we 
have chosen $T_{\rm kin}$ = 6, 10 and 14 K. Neutral carbon column densities
are adjusted to yield, at the relevant kinetic temperature, the observed 
[CI]/$\co$ (2--1) ratio of 0.24. By combining these with the model 
calculations of van Dishoeck $\&$ Black (1988), we then estimate the 
gas-phase carbon fraction $N_{\rm C}/N_{\rm H}$ = $(N(CO) + 
N(C))/(2N(H_{2}$ + N(HI)). From Bajaja $\&$ Shane (1982) and Brinks 
$\&$ Shane (1984) we find a neutral hydrogen column density $N(HI)$ = 
1.8 $\times$ 10$^{20}$ $\cm2$ at the position of D~478. 

Parameters thus determined for single-component models are shown in Table 
2, together with corresponding model line ratios. Although these models 
generally yield greater differences between the $\co$ (2-1)/(1-0) and 
(3-2)/(2-1) ratios than observed, the 14 K model fits the observations
within the errors. The 10 K model is barely consistent and the 6 K 
model fails to reproduce the observed ratios, predicting too low a 
$\co$ $J$ = 3--2 strength. Loinard $\&$ Allen (1998) have argued that 
the $\co$ structure of D~478 can be explained by a large, 
dense and very smooth cloud in LTE at even lower temperatures close to 
that of the cosmic background. Although it is possible to reproduce the 
$\co$ line ratios in this manner, the observed $\13co$ (2--1)/(1--0) ratio 
of about 0.5 is only obtained at high $\13co$ optical depths requiring 
$\co$/$\13co$ ratios of order unity, which is clearly not the case. 
Alternatively, the observed $\co$/$\13co$ ratios can be reproduced with 
lower $\13co$ optical depths satisfying one of the two observed $\co$/$\13co$ 
ratios, but the $\13co$ (2--1)/(1--0) ratio then is only a third of the 
actually observed value.

For such reasons, Allen et al. (1995) concluded to the necessity of a model
incorporating at least two components. In their analysis, they modelled the CO
emission from D~478 with two different density/temperature components, and
showed that the observed line ratios are, for instance, reproduced by a cold 
low-density gas ($T_{\rm k} \approx$ = 4 K, $n_{\rm H}$ = 100 $\cc$, d$V$ = 
5 $\kms$) and a high-density gas ($T_{\rm k}$ = 7 K, $n_{\rm H}$ = 3000 $\cc$, 
d$V$ = 10 $\kms$), both with a low surface-filling factor.

We have also calculated two-component models. As there are eleven
adjustable parameters (six column densities, two densities, two
temperatures and the relative contributions of the two components
to the total emission), no unique solution can be obtained. We have chosen
to model the same $\h2$ densities as given in the example by Allen et al. 
(1995) and similar $\co$ column densities, and calculate the remaining
parameters as a function of temperature. Furthermore, we require the
column density ratio of the tenuous and the dense components to be the
same for $\13co$ and for $\co$. Over the range of applicable parameters,
the [CI] emission from the dense component is more efficiently produced 
(i.e. characterized by a higher ratio of [CI] intensity to carbon column 
density) than that from the tenuous component. By taking the maximum 
permissible contribution from the dense component, and ascribing the
remainder to the tenuous component, we have minimized overall neutral 
carbon column densities required, and thus maximized permitted total
hydrogen column densities.  Although other combinations of input
parameters are possible, the range of admissible parameters is nevertheless
limited by the observations. For instance, changes in the assumed $\h2$ 
densities and $\co$ column densities, immediately require adjustment of 
the relative weights of the two components in order to reproduce the same 
line ratios which tends to minimize the effect of these changes on the
parameter ratios as well.

In Table 2, we present four such cases, for temperatures varying from 6 to 
14 K. As an example, the 6/6 K case corresponds closest to the example 
given by Allen et al. (1995), with the surface filling factor of the 
low-density component twice that of the high-density component. For this 
case, we obtain a reasonable fit for a $\co$/$\13co$ column-density ratio 
of 360. If, instead, we require the column density ratio to be 90, $\13co$ 
optical depths increase significantly and we obtain for the same relative 
filling factors a $\co$/$\13co$ {\it temperature} ratio of less than 4, 
inconsistent with the observed values. This may be repaired by increasing 
the relative filling factors to 1:6 or higher, but then the $\co$ 
$J$=2--1/$J$=1--0 and especially the $J$=3--2/$J$=2--1 ratios drop to 
unacceptably low values of 0.32 and 0.19 respectively. 

As Table 2 shows, increasing the temperature generally leads to higher 
$\co/\13co$ ratios, lower ratios of neutral carbon to CO column density
and a lower required contribution of the dense component. The dense 
component has low $N(C)/N(CO)$ ratios characteristic of Giant Molecular 
Cloud complexes in the Milky Way (0.1--0.2: Keene et al. 1985), whereas 
the tenuous component exhibits ratios similar to those found in Galactic 
translucent and dark clouds (0.3--3: Stark $\&$ van Dishoeck 
1994; Stark et al. 1996).

If we increase the temperature of the dense component over that of the 
tenuous component, the $\13co$ column densities change only by a small 
amount, the implied C column densities decrease and the relative 
contribution of the tenuous component also decreases. The predicted line 
ratios, although somewhat different, are consistent with the observed 
ratios. Thus, from a radiative-transfer point of view, the temperature of 
D~478 is not strongly constrained in a two-component model, although we 
note that the higher temperature (10/10 K and 10/14 K) models give slightly 
better fits to the observed line ratios. In the models in Table 2, the 
[CI] emission is optically thick at 6 K, and optically thin at 14 K. In all 
cases, the expected strength of the 809 GHz line is (much) less than 10 
per cent of that of the 492 GHz line. As neutral carbon radiates 
ineffectively at temperatures below 10 K, the observed [CI] intensity rapidly 
requires very large C column densities and high $N(C)/N(CO)$ ratios if 
lower temperatures are assumed. This again rules out satisfactory
solutions at very low temperatures. 

\begin{table*}
\caption[]{Beam-averaged parameters for D~478}
\begin{flushleft}
\begin{tabular}{llccccccc}
\hline
\noalign{\smallskip} 
   & Units	& \multicolumn{3}{c}{Single-Component Models}  
                & \multicolumn{4}{c}{Two-Component Models} \\
\noalign{\smallskip}
\hline
\noalign{\smallskip}
$T_{\rm kin}$   & (K)               &  6   & 10    & 14    & 6, 6  & 6, 10 & 10, 10 & 10, 14 \\
\noalign{\smallskip}
$N(\co)$	& $(10^{16} \cm2)$  &  5   &  4    &  4    & 160   & 135   & 50     & 45  \\
$N(\rm C)$	& $(10^{16} \cm2)$  & 26   &  6    &  4    & 130   &  35   & 25     & 15  \\
$N(\rm C)/N(\rm CO)$  & 	    &  5   & 1.5   &  1    & 0.8   & 0.3   & 0.5    & 0.3 \\
$N_{\rm C}$	& $(10^{16} \cm2)$  & 31   & 10    &  8    & 290   & 170   & 75     & 60  \\
\noalign{\smallskip}
\hline
\noalign{\smallskip}
$N_{\rm H}$	& $(10^{21} \cm2)$  & 1.6  &  1.7  &  1.2  &   3.5 &   4   &   2    &  2   \\
$N_{\rm C}/N_{\rm H}$ & $(10^{-4})$ &  2   &  0.6  &  0.7  &   8   &   4.5 &  3.5   &  3   \\
$N(\h2)$	& $(10^{21} \cm2)$  & 0.7  &  0.7  &  0.5  &   1.7 &  1.8  &  1.0   & 0.9  \\
$f(\h2)$ (dense) &		    & 0.22 & 0.11  &  0.10 &  0.27 &  0.27 &  0.06  & 0.06 \\
$f(\h2)$ (tenuous) &		    & ---  & ---   &  ---  &  0.54 &  0.27 &  0.33  & 0.24 \\
$N(\co)/N(\h2)$ & $(10^{-4})$       & 0.7  &  0.5  &  0.8  &   9.5 &  7.5  &   5    &  5   \\
d$M$/d$A$ & (M$_{\odot}$\,pc$^{-2})$ &  4  &   4   &   3   &    9  &   9   &   5    &  5    \\
$X$ & $(10^{20}\,\cm2\,(\kkms)^{-1})$ & 0.5 &  0.5 &  0.3  &   1.1 &  1.2  &  0.7   & 0.6   \\
\noalign{\smallskip}
\hline
\end{tabular}
\end{flushleft}
\end{table*}

\section{Discussion}

\subsection{The environment of D~478}

We have used the chemical models by van Dishoeck $\&$ Black (1988)
assuming an incident radiation field $I_{\rm UV}$ $\approx$ 0.5,
corresponding to $I_{\rm 1000}$ = 2.25 $\times$ 10$^{-8}$ photons 
s$^{-1}$ cm$^{-2}$. Is this a realistic value? No measurements at 
$\lambda$ = 100 nm exist for M 31, but we may estimate the value of 
$I_{\rm UV}$ at the location D~478 by using data obtained at 155 nm
(Carruthers et al. 1978; Wu et al. 1980; Israel et al. 1986). After 
a correction of $E(B-V)$ = 0.11 for Galactic foreground extinction 
only, we find from these data an irradiation of D~478 by about
$I_{\rm 1550}$ = 2.5 $\times$ 10$^{-8}$ photons s$^{-1}$ cm$^{-2}$. A
similar result was obtained independently by Koper (1993). About 60$\%$ 
of this originates in the relatively nearby M~31 bulge, while about 40$\%$ 
is contributed by the more distant star-forming ring. A somewhat 
uncertain extrapolation of the 155 nm fluxes to $\lambda$ = 100 nm
then suggests irradiation by $I_{\rm UV}$ $\approx$ 0.3--0.4, i.e
close to the assumed value. A substantially lower value of $I_{\rm UV}$
$\approx$ 0.1 would change our conclusions only slightly: it would require
lower $\h2$ column densities, and somewhat higher carbon abundances.
In the absence of measurements, we have neglected the possible presence
of ionized carbon (C$^{+}$). If C$^{+}$ is present in any significant
amount, it would raise the total carbon column $N_{\rm C}$ and lower
the fraction of all carbon contained in CO. In turn, this would tend
to {\it lower} the resulting $\h2$ column densities, and lead to
significantly higher carbon abundances. 

The models would be better constrained if we had an independent way of
estimating total hydrogen or molecular hydrogen column densities. This
would be the case if, for instance, the CO intensity to $\h2$ column 
density conversion factor $X$ or the carbon abundance in the center of M~31 
were known. Unfortunately, neither is well-determined. In the disk of our 
Galaxy, a conversion factor of $X$ = 2 $\times$ 10$^{20}$ mol $\cm2$ 
$(\kkms)^{-1}$ (Strong $\&$ Mattox 1996) applies, within a factor of two. 
However, in the central part of the Galaxy $X$ is a factor of 3--10 lower 
(Sodroski et al. 1995), whereas in low-metallicity environments $X$ may be 
up by one or even two orders of magnitude (Israel 1997). 

Oxygen abundances in HII regions and supernova remnants in the disk of 
M~31, studied by Blair et al. (1982), indicate the presence of a gradient, 
whose extrapolation inwards yields an [O]/[H] abundance of about 1.3 $\times$ 
10$^{-3}$ at the galactocentric radius of D~478. Data summarized by Garnett 
et al. (1995) and Kobulnicky $\&$ Skillman (1998) suggest that at such high
oxygen abundances, the ratio of carbon to oxygen abundances is given
by  log [C]/[O] = $-0.2\,\pm\,0.3$ so that [C]/[H] = $8 (+8,-4) \times 
10^{-4}$. This result has a large uncertainty, however, because no direct 
abundance measurements exist within $R$ = 5 kpc, because the M~31 disk 
data exhibit a relatively large scatter, and because the relation between 
[O]/[H] and [C]/[O] abundances is not well-established at such high [O]/[H] 
abundances. In addition, a significant fraction of all carbon in quiescent 
dark clouds such as D~478 is expected to be depleted onto dust grains, 
leaving actual gas-phase carbon abundances substantially lower, by factors 
of order $\delta_{\rm C} \approx$ 0.4. This would lead us to expect total
column density ratios $N_{\rm C}/N_{\rm H} \approx 3 \times 10^{-4}$.

Moreover, an inward extrapolation from the disk may not be representative
of conditions in D~478. The velocity of D~478 is anomalously low for its
projected distance to the nucleus. In position-velocity maps presented
by Brinks $\&$ Shane (1984), one of which is reproduced by Loinard et al. 
(1995), it coincides with a loop-like HI structure. This 
structure exhibits a north-south symmetry with respect to the center, but 
is most pronounced north of the center. It appears to represent rapidly 
rotating gas in the inner part of M~31, moving in inclined elliptical 
orbits in a weak bar-like potential (cf. Stark $\&$ Binney 1994). D~478 
is located close to the northern tangential point of this rotating structure. 
It is conceivable that this is the signature of a merging event in which a 
small late-type galaxy fell into the large spiral M~31. Such a notion is 
supported by the presence of a double nucleus in M~31 (cf. Lauer et al. 1993).
In that case, the metallicity of the inner gas clouds
may be substantially lower than expected from extrapolation of the
disk gradient. If the metallicity is comparable to that of the outer
disk of M~31 we have [O]/[H] = 4 $\times$ 10$^{-4}$ and again following
Garnett et al. (1995) we find a low total carbon abundance [C]/[H] 
= 1.5 $\times$ 10$^{-4}$. Taking into account depletion, this case
would lead us to expect $N_{\rm C}/N_{\rm H} \approx 0.7 \times 10^{-4}$.

\subsection{Conditions in D~478}

In Table 3, we have used the ratio of the observed to model line intensities
to calculate {\it beam-averaged} column densities of carbon monoxide and
neutral carbon, as well as those of total carbon (C + CO), total hydrogen,
and molecular hydrogen. We have used the total hydrogen column densities, 
multiplied by 1.35 to take into account a contribution by helium, to 
calculate the projected mass density d$M$/d$A$ (corrected by cos\,$i$ 
with $i$ = 77$^{\circ}$) and we also determined the CO -- $\h2$ conversion 
factor $X$. As shown in the preceding section, we have, if anything, 
overestimated hydrogen column densities (and masses), and underestimated 
carbon abundances. In addition, we have estimated surface filling factors 
$f(\h2)$ by dividing the observed velocity-integrated line intensities by 
the model intensities, and assuming that the surfaces covered by CO and 
by $\h2$ are identical.

First we consider the single-density models. The physical equivalent
of these models is that of high-density filamentary clouds rather
uniformly but incompletely filling a large surface area. We have already 
ruled out the model for $T_{\rm k}$ = 6 K, because it does not reproduce the
observed $\co$ $J$=3--2 intensity. Both the 10 K and 14 K models fit the
observed intensities although the predicted $\co$ $J$=3--2 intensity is 
low compared to the observed value (Table 2), rendering the 10 K model in
particular somewhat marginal. However, single-component models are 
consistent only with {\sl low} [C]/[H] ratios. They are ruled out if we 
accept the extrapolated high carbon abundances, but would be just consistent 
with an anomalously low metallicity as discussed above. The single-density 
models furthermore require rather low $\co/\13co$ abundance ratios of about 
20. This is not out of the question, because in cold diffuse and translucent 
clouds, such as described by the single-density models in Table 2, 
ion-molecule exchange reactions may favour the formation of $\13co$ at the 
expense of $\co$ for temperatures $T_{\rm k} < 36$K although $\13co$ remains 
more susceptible to photodissociation than $\co$. Surface filling factors 
are between 0.2 and 0.1, the lower values pertaining to the higher 
temperatures. The relatively low $\h2$ column densities imply 
projected mass densities of only about 4 M$_{\odot}$\,pc$^{-2}$ (Table 3),
and $X$ values about five times lower than those in the Solar
Neighbourhood. We conclude that for D~478 single-component models at 
temperatures $T_{\rm kin}$ $\geq$ 10 K on the one hand cannot be ruled 
out, but on the other hand do not appear to be very convincing.

The two-density models provide better fits to the observations; like the 
single-component models they have average $\h2$ column densities below
the star formation treshold in keeping with the quiescent appearance
of D~478. The 6/6 K model requires a relatively large $N_{\rm C}/N_{\rm H}$ 
ratio of eight, implying a total carbon abundance [C]/[H] $\geq$ 2 $\times$
10$^{-3}$ which seems rather high. More reasonable carbon abundances
would imply higher $\h2$ column densities and projected gas mass densities,
but the large neutral carbon column densities required by the observed line 
strength at the assumed temperature of 6 K would then imply $N(C)/N(CO)$ 
ratios much larger than allowed at such column densities. 

With increasing temperature, the CO and [CI] column densities required by
the observed line emission become more modest. Because the carbon/hydrogen
column density ratio also drops, total hydrogen and molecular hydrogen
column densities change relatively little. At these temperatures, 
$N_{\rm C}/N_{\rm H} \approx 4 \times 10^{-4}$, which corresponds to 
a total carbon abundance [C]/[H] $\approx 1 \times 10^{-3}$ if
$\delta_{\rm C}$ = 0.4. This is about 2.5 times the value ascribed to the 
Solar Neighbourhood, and close to the value expected from the extrapolation 
of M~31 disk abundances. The CO intensity to $\h2$ column density value for 
these models is relatively well-determined with $X$ = 0.9$\pm$0.3 $\times 
10^{20}$ mol cm$^{-2}$ $(\kkms)^{-1}$, or about half of the Solar 
Neighbourhood value. Projected mass densities are in the range of 
5--9 M$_{\odot}$ pc$^{-2}$. Surface filling factors for the emission
within the beam are given in Table 3 for both the high-density and the
low-density components; the ratio of the two is for each model, of
course, equal to the ratio of the weights given in Table 2. Again,
the lower temperature models have the largest filling factors. The two
models with dense gas temperatures of 10 K yield surface filling
factors of 6$\%$ for the dense gas and about 30$\%$ for the tenuous gas.
These values appear to be consistent with the small fraction (15$\%$)
of the single-dish CO flux recovered in the interferometric map by
Loinard $\&$ Allen (1998). The model $\h2$ column densities divided by 
the input $\h2$ densities provides a measure for the line-of-sight
depth z through the clouds in the emitting surface fraction. For the
two-component models in Table 3, we find 0.3 pc < $z_{d}$ < 0.7 pc and 
4.5 pc < $z_{t}$ < 7.5 pc. The subscripts $d$ and $t$ refer to the
dense and the tenuous components respectively, and the larger depths
occur at the higher temperatures. Both the surface filling factors and 
the subcloud depths suggest that the structure of D~478 is highly
fragmented and filamentary.
 
The models discussed do not fully constrain the physical conditions
applying to D~478. Various other combinations of parameters could
be used as input. Multi-density models incorporating temperature
gradients might be better representations of the actual structure
of D~478, and could easily be made to satisfy the observations.
For instance, adding a contribution of material at intermediate
densities and temperatures to the two-component models in Table 2
within the limits imposed by observed line ratios would indeed
change the detailed set of physical parameters, but would not 
substantially change the overall result. We conclude
that the observations of D~478, and in particular the relatively
strong [CI] emission appear to rule out the presence of very large
amounts of molecular gas at very low temperatures in this complex.
At the very least, our models show that there is no compelling need
to assume such a state of affairs. The interferometric results 
obtained by Loinard $\&$ Allen (1998) indicate a relatively small
contribution by dense, compact material. This suggests that the
10/10 K and 10/14 K models with their small surface-filling
fraction of the dense component are a better representation than
the lower-temperature models. 

Most likely, D~478 is rather similar to dark cloud complexes in 
our own Galaxy, at the higher metallicity consistent with its 
central location in M~31. It has a modest projected mass density 
and a relatively low $X$-factor, both of which are consistent with 
this conclusion. A comparison with the Galactic Taurus-Auriga dark cloud 
complex at a distance of 140 pc is illustrative. We have already noted
similar [CI]/$^{13}$CO (2-1) ratios (Schilke et al. 1995). Likewise,
the Taurus $^{12}$CO (2--1)/(1--0) ratio of 0.53 (Sakamoto et al. 1997)
is close to that of D~478. More in detail, an area 30$^{\circ}$ (80 pc) 
across was mapped in CO by Ungerechts $\&$ Thaddeus (1987). Ignoring the 
more distant and unrelated clouds, we find that the mean integrated line 
intensity of the complex is 4.5 $\kkms$ with individual clouds having 
intensities between 3 and 15 $\kkms$. The surface filling factor of the 
complex is 0.3, and its appearance is indeed very fragmented and clumpy.
The complex has a mean mass-density d$M$/d$A$ = 7 M$_{\odot}$\,pc$^{-2}$. 
Both the observed line ratios and the derived quantities thus suggest that
D~478 is similar to a larger version of the Taurus-Auriga dark cloud
complex.

Assuming an overall cloud size of 1$'$ (Allen
$\&$ Lequeux 1993) and including a contribution by helium, the
gas mass of D~478 is of the order of 1 $\times$ 10$^{6}$ M$_{\odot}$. 
This is an order of magnitude lower than the virial mass deduced by 
Allen $\&$ Lequeux (1993) and Loinard $\&$ Allen (1998). However, 
it is doubtful whether molecular cloud complexes are virialized at 
all. Moreover, the shape of the line profiles of D~478 suggests the 
presence of at least two distinct components. Its location near the 
tangential point of a rapidly rotating structure with noncircular 
motions suggests that various individual clouds, separated by significant
distances, may be seen with very similar velocities in the same
line of sight; the appearance of the HI distribution in the
position-velocity maps published by Brinks $\&$ Shane (1984) lends
credibility to this suggestion. In this respect it is relevant to note
that the velocity width of line spectra incorporating emission of 
various unrelated clouds in the same beam will reflect the cloud-cloud
velocity dispersion of about 9 $\kms$ (Stark 1979) in addition to
any systematic velocity shifts and the velocity widths of the individual
clouds. Applying the virial equation to such a collection of clouds
will lead to a potentially large overestimate of the actual mass.

\section{Conclusions}

\begin{enumerate}
\item We have detected the 492 GHz emission line from the $\pci$
transition towards the dark cloud D~478 in M~31. The [CI] line strength
is comparable to that of $\13co$ $J$=1--0 and about twice as strong
as that of the $\13co$ $J$=2--1 transition. This is similar to the
intensity ratios found in Galactic dark and translucent clouds.

\item We have used $\co$ and $\13co$ intensities from the literature
together with the newly determined [CI] intensity to model conditions
in the D~478 cloud complex. We have considered both single-density
filamentary clouds and two-density core/envelope clouds. In particular, 
inhomogeneous (two-component) models produce good fits to the observations.
Dense core $N(C)/N(CO)$ ratios are found to be similar to those of Giant 
Molecular Clouds in the Milky Way, whereas the ratios derived for the
extended more tenous gas are similar to those found in Galactic dark and
translucent clouds. 

\item D~478 appears to be similar to Galactic dark cloud complexes, such as
the Taurus-Auriga dark cloud complex. Most likely, it is characterized by a 
higher metallicity consistent with an extrapolation of M~31 disk abundances 
to its central 
location in the galaxy. Its kinetic temperature is of the order of 10 K.
There appears to be no need to assign very low temperatures and very high 
$\h2$ mass densities to the D~478 complex; our results do not support the 
virial mass surface densities of D~478 of $\sim$100 M$_{\odot}$\,pc$^{-2}$ 
suggested by Loinard $\&$ Allen (1998).

\item Actual cloud mass densities projected onto the M~31 midplane are of 
the order of 5--10 M$_{\odot}$\,pc$^{-2}$. The CO to $\h2$ conversion factor 
is $X$ = 0.9$\,\pm\,$0.3 $\times$ 10$^{20}$ cm$^{-2}$ $(\kkms)^{-1}$, about 
half the value in the Solar Neighbourhood.

\item The D~478 complex may not be a single entity, but instead
may consist of various clouds at different locations projected
onto the same line-of-sight towards the tangential direction
of a rapidly rotating structure characterized by noncircular
velocities.

\end{enumerate}
\acknowledgements

We are indebted to Ewine van Dishoeck and David Jansen from the Leiden 
astrochemistry group for letting us use their statistical equilibrium
calculation models and benefitted from discussions with Ewine van Dishoeck.
We also thank James Lequeux for kindly providing us with the unpublished 
full-resolution parameters of the $\co$ $J$=2--1 profile of D~478.
Critical remarks on an earlier draft by Ron Allen and Lasurent Loinard 
led to a substantial improvement in the paper.

\end{document}